\documentclass[12pt,a4]{article}

\usepackage[T1]{fontenc}
\usepackage{authblk}

\usepackage{amssymb}
\usepackage{amsmath}
\usepackage{graphicx}

\usepackage{algorithm}
\usepackage{algorithmicx}
\usepackage{algpseudocode}

\newcommand{\INDSTATE}[1][1]{\State\hspace{#1\algorithmicindent}}

\newcommand\Ham{\hat{H}}
\newcommand\nr{{\it n}({\bf r})}
\newcommand\kv{{\bf k}}
\newcommand\rv{{\bf r}}
\newcommand\hsumpart[2]{\left(#1_{\it L'}^{a,{\bf
        G'}}\right)^{\ast}\thinspace T_{{\it L'},{\it
      L};a}^{\left[#1#2\right]}\thinspace#2_{\it L}^{a,{\bf G}}}

\begin{document}

\title{Accelerating the computation of FLAPW methods on heterogeneous architectures}

\author[2]{Davor Davidovi\' c}
\author[1]{Diego Fabregat-Traver}
\author[1]{Markus H\"ohnerbach}
\author[1,3]{Edoardo di Napoli}

\affil[1]{Aachen Institute for Advanced Study in Computational Engineering Science, RWTH~Aachen, 52062 Aachen, Germany}

\affil[2]{Ru\dj{}er Bo\v{s}kovi\'c Institute, 10000 Zagreb, Croatia}

\affil[3]{J\"ulich Supercomuting Center and JARA,J\"ulich, Germany}

\maketitle

\begin{abstract}
	Legacy codes in computational science and engineering have been very 
	successful in providing essential functionality to researchers. However, 
	they are not capable of exploiting the massive parallelism provided by 
	emerging heterogeneous architectures.       
    The lack of portable performance and scalability puts them
    at high risk: either they evolve or they are doomed to disappear.
    One example of legacy code which would heavily benefit from a modern design
    is FLEUR, a software for electronic structure calculations. In previous
    work, the computational bottleneck of FLEUR was partially re-engineered to
    have a modular design that relies on standard building blocks, namely BLAS
    and LAPACK.
    In this paper, we demonstrate how the initial redesign enables the
    portability to heterogeneous architectures. More specifically,
    we study different approaches to port the code to architectures consisting
    of multi-core CPUs equipped with one or more coprocessors such as Nvidia GPUs and
    Intel Xeon Phis. Our final code attains over 70\% of the architectures'
    peak performance, and outperforms Nvidia's and Intel's libraries.
    Finally, on JURECA, the supercomputer where FLEUR is often executed,
    the code takes advantage of the full power of the computing nodes, attaining
    $5\times$ speedup over the sole use of the CPUs.

\end{abstract}

\newpage
\pagenumbering{arabic}

\section{Introduction} \label{sec:intro}

Modern software is typically developed with modularity, portability
and scalability in mind, while older codes often rely on a direct
implementation of mathematical formulas. The latter approach limits
the reuse of the code on emerging computing architectures and constitutes 
a substantial portability challenge:
each of these formulas must be rewritten and optimized for each new
architecture. One such example is FLEUR, a software for electronic
structure calculations developed during the last three decades at the
Forschungszentrum J\"ulich (Germany)~\cite{FLEUR}.

Today, in a context where massively-parallel heterogeneous
architectures have become ubiquitous, legacy code must be
re-engineered and adapted to make an efficient use of these
architectures.  A critical insight in writing long-lasting scientific
code is to have a modular design where, at the bottom layers, the
computational bottlenecks are written in terms of kernels.  Examples
of such kernels are fast Fourier transforms, matrix products, and
eigensolvers, provided by a number of commercial as well as academic
libraries, such as Intel MKL~\cite{IntelMKL}, MAGMA~\cite{Magma},
cuBLAS~\cite{CUBLAS}, cuFFT, and
ELPA~\cite{AuckenthalerElpa,marek2014elpa}.  Over time, standardized
libraries implement highly-optimized routines for the most critical
kernels on various computing architectures, from which applications
can benefit automatically, without the need to change the code.

The most outstanding example is the Basic Linear Algebra Subproblems
(BLAS) library. In reality, the BLAS is an interface that defines a
number of common building blocks (including matrix products and linear
systems), which was born after the realization that standardization
was critical to increase productivity and portability. Today, the BLAS
is the first library ported and optimized for each new architecture,
and is used as a building block for many other higher-level
libraries. Therefore, writing software on top of the BLAS is a guarantee
for performance portability and scalability.

In order to demonstrate the benefits of modern engineering and
portability, Di~Napoli et al. underwent a major effort to rework one
of the computational bottlenecks of FLEUR~\cite{DiNapoli:HSDLA}: the
construction of the so-called Hamiltonian and Overlap matrices. The
concepts of abstraction and encapsulation were lacking in the original
code, and the different modules were tightly coupled.  Therefore, the
authors rewrote the code from scratch. At this point, they decided to
analyze the mathematics behind the code and rewrite the formulas at a
higher level, resulting in a series of matrix operations. These matrix
operations were then cast to kernels provided by the BLAS and LAPACK
libraries.  The new algorithm (HSDLA) attains speedups ranging from
1.5$\times$ to 2.5$\times$ with respect to the corresponding FLEUR
code on multi-core CPUs. More importantly, the new code enables
performance portability with limited programming effort which was
demonstrated in our previous work~\cite{jhpcs/Fabregat}. We analyzed
the performance portability of the HSDLA code on heterogeneous
architectures consisting of multi-core shared-memory CPUs and one or
more Graphical Processing Units (GPUs).  In particular, we showed that
the HSDLA code can be ported to emerging architectures with minor or
relatively straightforward code changes by employing highly-optimized
libraries such as NVIDIA's CUBLASXT or Intel's MKL. However, this
approach attained only up to $69$\% of the theoretical system's peak
performance, thus still significantly underutilizing the available
resources.

In this paper we close the gaps still open in the previous work to
demonstrate that the initial effort striving for modularity and
portability indeed facilitates quick performance portability to
complex heterogeneous architectures comprising multi-core
shared-memory CPUs and either GPUs or Phi coprocessors.  We first
attempt a (almost) plug-and-play solution and quantify the benefits
and limitations to make the best of these architectures. Then, we
study alternative approaches to leverage existing optimized BLAS
kernels for each component to obtain efficient hybrid
routines. Finally, we identify scalability bottlenecks in sequential
or poorly-scalable sections of the algorithm and propose solutions to
limit their impact in the overall scalability.  As our experimental
results obtained on a number of architectures illustrate, the
resulting code makes an efficient use of the computing resources and
provides the expected boost to the application code.

\paragraph{Contributions}
We make the following key contributions. First, we perform a portability study
of the code, once it has been rewritten in terms of a standardized interface, and
point out the remaining issues for a successful port to heterogeneous architectures. 
We then analyze different approaches to overcome these limitations, implement them and
quantify the improvements. Finally, we contribute a high-performance and scalable
implementation to compute the Hamiltonian and Overlap matrices, a computational bottleneck
in FLAPW methods.

\paragraph{Organization of the paper}
The remainder of the paper is organized as follows. Section~\ref{sec:flapw}
provides the background on Density Functional Theory and the math behind the computation to generate the Hamiltonian and Overlap matrices. Section~\ref{sec:hsdla} gives an overview of the HSDLA algorithm for the generation of these matrices and Section~\ref{sec:relwork} provides overview of the related work. The improvements of the HSDLA algorithm as well as our implementation on the heterogeneous architectures, are described in Section~\ref{sec:hdslaHetero}.
Section~\ref{sec:experiments} presents experimental results for a collection of
test cases and hybrid architectures. Finally, Sec.~\ref{sec:conclusions} draws
conclusion and discusses future research directions.

\section{Computing the Hamiltonian and Overlap matrices in FLEUR}
\label{sec:flapw}

FLEUR is a scientific computing code, well-known in the community of
condensed matter physicists, for calculating ground-state and
excited-state properties of solids. This is one of few Density
Functional Theory (DFT) which computes the electron structure of an
atomic compound using the Full-potential Linearized Augmented Plane
Wave (FLAPW) discretization approach. In this section we give a brief introduction
of the fundamental mathematical aspects of DFT, and its
practical realization by the FLAPW method. The material presented does
not require a special background knowledge in quantum physics and allows the
non-specialist to understand the origin of one of the most
computational intensive parts of the FLEUR code, 
the construction of the Hamiltonian and Overlap matrices.


\subsection{DFT and the FLAPW method}

In the last two decades, Density Functional Theory
(DFT)~\cite{Nogueira:1391332,Sholl:2011td} has become the ``standard
model'' in Materials Science. Within the DFT framework, it is possible
to simulate the physical properties of complex quantum mechanical
systems made of few dozens up to few hundreds of atoms. The core of
the method relies on the simultaneous solution of a set of partial
differential equations. These equations are determined by a
Hamiltonian operator $\Ham$ containing a so-called effective potential
$V_0[n]$, where $\nr$ is a function of the radial coordinate ${\bf r}$
known as the one-particle electron density. 

In turn, the solutions of the equations $\psi_i({\bf r})$ determine
the one-particle electron density $\nr$ used in calculating the
effective potential $V_0$. Because of the dependence of the
Hamiltonian on the set of $\psi_i({\bf r})$ through $\nr$, the
governing equations are highly non-linear and cannot be solved
directly.
\begin{eqnarray}
\label{eq:KSeq}
\begin{array}[l]{l}
		\Ham \psi_i({\bf r}) = \left( -\frac{\hbar^2}{2m} \nabla^2 + V_0({\bf r}) \right) \psi_i({\bf r}) = \epsilon_i \psi_i({\bf r}) \quad ; \quad \epsilon_1 \leq \epsilon_2 \leq \dots\\[2mm]
		\nr = \sum_i^N f_i |\psi_i({\bf r})|^2
\end{array}
\end{eqnarray}

In practice, the equations above, also known as Kohn-Sham
(KS)~\cite{Kohn:1965zzb}, are solved by self-consistent iteration; an initial guess for
$n_0({\bf r})$ is used to calculate the effective potential $V_0$ which, in
turn, is inserted in Eq.~\eqref{eq:KSeq} whose solutions,
$\psi_i({\bf r})$, are used to compute a new charge density
$n_1({\bf r})$. Convergence is checked by comparing the new density to the
starting one. When convergence is not reached, an opportune mixing of
the two densities is selected as a new guess, and the process is
repeated. 

\sloppypar
In principle, the theory only requires as input the quantum numbers
and the positions of the atoms that are part of the investigated
system. In practice, there is a plethora of DFT methods which depends
on the {\em discretization} used to parameterize the KS equations. The
discretization in the Full-potential Linearized Augmented Plane Wave
(FLAPW) method~\cite{Wimmer:1981hd,Jansen:1984bc} is based on a plane wave expansion of
$\psi_{\kv,\nu}(\rv)$, where the momentum vector $\kv$ and the band
index $\nu$ replace the generic index $i$. The \kv-point wave function
$\psi_{\kv,\nu}(\rv) = \sum_{|{\bf G + k}|\leq {\bf K}_{max}} c^{\bf
  G}_{\kv,\nu} \varphi_{\bf G}(\kv,\rv)$
is expanded in terms of a finite basis set $\varphi_{\bf G}(\kv,\rv)$ indexed
by the vectors ${\bf G}$ lying in the lattice reciprocal to
configuration space up to a chosen cut-off value ${\bf K}_{max}$. In
FLAPW, the physical (configuration) space of the simulation cell is
divided into spherical regions, called Muffin-Tin (MT) spheres,
centered around atomic nuclei, and interstitial areas between the MT
spheres. The basis set $\varphi_{\bf G}(\kv,\rv)$ takes a different
expression depending on the region
\begin{eqnarray}
\label{eq:basis}
\varphi_{\bf G}(\kv,\rv) \propto \left\{
	\begin{array}[l]{lr}
	e^{i({\bf k+G})\rv} & \quad \textrm{\small Interstitial}\\
	\displaystyle\sum_{\it l,m} \left[A^{a,{\bf G}}_{\it l,m}(\kv) u^{a}_{\it l}(r) 
	+ B^{a,{\bf G}}_{\it l,m}(\kv) \dot{u}^{a}_{\it l}(r)
          \right] Y_{\it l,m}(\hat{\bf r}_{a}) & \quad
        \textrm{\small $a^{th}$ Muffin Tin}\\
	\end{array}
\right.
\end{eqnarray}
where the coefficients $A^{a,{\bf G}}_{\it l,m}(\kv)$ and
$B^{a,{\bf G}}_{\it l,m}(\kv)$ are determined by imposing
continuity of $\varphi_{\bf G}(\kv,\rv)$ and its derivative at the
boundary of the MTs. Due to this expansion the KS equations naturally
translate to a set of generalized eigenvalue problems
$\sum_{\bf G'} \left[ H_{\bf G,G'}(\kv) - \lambda_{\kv\nu} S_{\bf
    G,G'}(\kv) \right] c^{\bf G'}_{\kv ,\nu} =0$
for the coefficients of the expansion $c^{\bf G'}_{\kv ,\nu}$ where the
Hamiltonian and Overlap matrices $H$ and $S$ are given by multiple
integrals and sums 
\begin{equation}
\label{eq:HSdef} 
\{H(\kv),S(\kv)\}_{\bf G,G'} = \sum_a \iint \varphi^{\ast}_{\bf
  G}(\kv,\rv) \{\Ham,I\} \varphi_{\bf G'}(\kv,\rv) {\rm d} \rv.
\end{equation}
Since the set of basis functions in Eq.~\eqref{eq:basis} is
implicitly labeled by the values the variable $\kv$ takes in the
Brillouin zone, there are multiple Hamiltonian and Overlap matrices,
one for each independent $\kv$-point.

\subsection{Building H and S}
Without loss of generality, we can abstract from the \kv -point index
and recover an explicit formulation of the $H_{\bf G,G'}$ and
$S_{\bf G,G'}$ matrices by substituting Eq.~\eqref{eq:basis} in
Eq.~\eqref{eq:HSdef} and carrying out the multiple integration. The
computation is particularly complex within the MT regions where the
initialization of the Hamiltonian and Overlap matrices is by far the
most computationally intensive task. By exploiting the properties of
the basis functions, the $H$ and $S$ matrices are directly expressed
as functions of the set of $A$ and $B$ coefficients.
\begin{equation}
\left(S\right)_{\bf G',G}
	=\sum_{a=1}^{N_A}\sum_{{\it l,m}}
     \left(A^{a,{\bf G'}}_{\it l,m}\right)^{\ast}A_{\it l,m}^{a,{\bf G}}
    +\left(B_{\it l,m}^{a,{\bf G'}}\right)^{\ast}B_{\it l,m}^{a,{\bf G}} \left\|
      \dot u_{\it l}^a\right\|^2
\label{eq:def_overlap}
\end{equation}
\begin{align}
\left(H\right)_{G',G}=\sum_{a=1}^{N_A}\sum_{L',L} & \left(\hsumpart AA\right)+\left(\hsumpart AB\right)\nonumber \\
+ & \left(\hsumpart BA\right)+\left(\hsumpart BB\right).
\label{eq:def_hamilton}
\end{align}
Notice that in Eq.~\eqref{eq:def_hamilton} for convenience we have
compacted the indexes ${\it l,m}$ into ${\it L}$, and expressed the
range of the index $a$ over all the distinct atom types $N_A$.  The
new matrices
$T_{L',L;a}^{\left[\dots\right]}\in \mathbb C^{N_L\times N_L}$ are
dense and their computation involves multiple integrals
between the basis functions and the non-spherical part of the
potential $V_0$ (See \cite[Appendix A.2]{DiNapoli:HSDLA} for details).
Due to the non-orthornormality of the basis function set
\eqref{eq:basis}, the matrix $S$ is non-diagonal, dense, and
generically positive definite with the exception of having few very
small singular values. On the opposite, $H$ is always non-definite and
both matrices are either complex Hermitian or real symmetric.

\section{The HSDLA algorithm} \label{sec:hsdla}

In the original FLEUR code the construction of matrices $H$ and $S$ was a direct 
implementation of mathematical formulae with the focus on the functionality rather 
than scalability. The performance scalability was significantly improved in the
HSDLA algorithm~\cite{DiNapoli:HSDLA}, in which the authors reformulated Eqs.~\eqref{eq:def_overlap} 
and~\eqref{eq:def_hamilton} in
terms of the coefficients $A$ and $B$. As a result, the entire construction
of matrices $H$ and $S$ is now cast in terms of matrix operands, as shown in Eqs.~\eqref{eq:h} and~\eqref{eq:s},

\begin{align}
    H &= \sum^{N_A}_{a=1} \underbrace{A^H_a T^{[AA]} A_a}_{H_{AA}} +
                          \underbrace{
                         A^H_a T^{[AB]} B_a + 
                         B^H_a T^{[BA]} A_a + 
                         B^H_a T^{[BB]} B_a}_{H_{AB+BA+BB}} 
    \label{eq:h}
                         \\
    S &= \sum^{N_A}_{a=1} A^H_a A_a + B^H_a U^H_a U_a B_a,
    \label{eq:s}
\end{align}

\noindent
where 
$A_a$ and $B_a \in \mathbb{C}^{N_L \times N_G}$,
$T^{[...]}_a \in \mathbb{C}^{N_L \times N_L}$,
$H$ and $S \in \mathbb{C}^{N_G \times N_G}$ are Hermitian, and
$U \in \mathbb{C}^{N_L \times N_L}$ is a diagonal matrix.
Typical values for the matrix sizes are 
$N_A \sim \mathcal{O}(100)$, 
$N_G \sim \mathcal{O}(1000)$ to $\mathcal{O}(10000)$, and 
$N_L \sim \mathcal{O}(100)$.

The HSDLA algorithm~\cite{DiNapoli:HSDLA} takes advantage of the
matrix formulation to cast the computation in terms of the highly optimized and
portable BLAS and LAPACK libraries. 
The algorithm is illustrated in Algorithm~\ref{alg:hands}. The main ideas behind it are:
1) exploiting the symmetries in the operations to reduce the computational cost, 
2) casting the computation in terms of efficient BLAS and LAPACK kernels, and
3) combining multiple operations on small matrices together to increase performance and scalability.
Although the algorithm exploits symmetry in matrices to reduce computational
costs, extra flops are introduced compared to the original version. However, these
are fast flops, run in highly-optimized BLAS-3 kernels, leading to significantly lower
execution time when compared to the original code.

\begin{algorithm}[!h]
\begin{algorithmic}[1]
    \State Backup $\hat{A} = A$, $\hat{B} = B$ \\
    // $H_{AB + BA + BB}$ 
    \For{$a := 1 \to N_A$}
        \State $Z_a = T^{[BA]}_a A_a$  \Comment{({\tt zgemm}: $8 N^2_L N_G$ Flops)}
        \State $Z_a = Z_a + \frac{1}{2} T^{[BB]}_a B_a$  \Comment{({\tt zhemm}: $8 N^2_L N_G$ Flops)}
        \State Stack $Z_a$ to $A$
        \State Stack $B_a$ to $B$
    \EndFor
    \State $H = A^H B + B^H A$         \Comment{({\tt zher2k}: $8 N_A N_L N^2_G$ Flops)}
    \\// S 
    \State Restore $A = \hat{A}$, $B = \hat{B}$
    \State $S = A^H A$      \Comment{({\tt zherk}: $4 N_A N_L N^2_G$ Flops)}
    \State $B = U B$        \Comment{({\tt scaling}: $2 N_A N_L N_G$ Flops)}
    \State $S = S + B^H B$  \Comment{({\tt zherk}: $4 N_A N_L N^2_G$ Flops)}\\
    // $H_{AA}$ 
    \For{$a := 1 \to N_A$}
        \State {\bf try:}
        \INDSTATE[0] $C_a = Cholesky(T^{[AA]}_a)$    \Comment{({\tt zpotrf}: $\frac{4}{3} N^3_L$ Flops)}
        \State {\bf success:}
        \INDSTATE[0] $Z_a = C^H_a A_a$               \Comment{({\tt ztrmm}: $4 N^2_L N_G$ Flops)}
        \INDSTATE[0] Stack $Z_a$ to $B_{T}$
        \State {\bf failure:}
        \INDSTATE[0] $Z_a = T^{[AA]}_a A_a$          \Comment{({\tt zhemm}: $8 N^2_L N_G$ Flops)}
        \INDSTATE[0] Stack $Z_a$ to $B_{B}$
        \INDSTATE[0] Stack $A_a$ to $A$
    \EndFor
    \State $H = H + B^H_{T} B_{T}$            \Comment{({\tt zherk}: $4 N_{A_{\text{HPD}}} N_L N^2_G$ Flops)}
    \State $H = H + A^H B_{B}$  \Comment{({\tt zgemm}: $8 N_{A_{\neg \text{HPD}}} N_L N^2_G$ Flops)}
    
\end{algorithmic}
    \caption{{\bf: Computation of the H and S matrices in HSDLA.}}
    \label{alg:hands}
\end{algorithm}

\begin{figure}[h]
    \centering
    \includegraphics[width=0.9\textwidth]{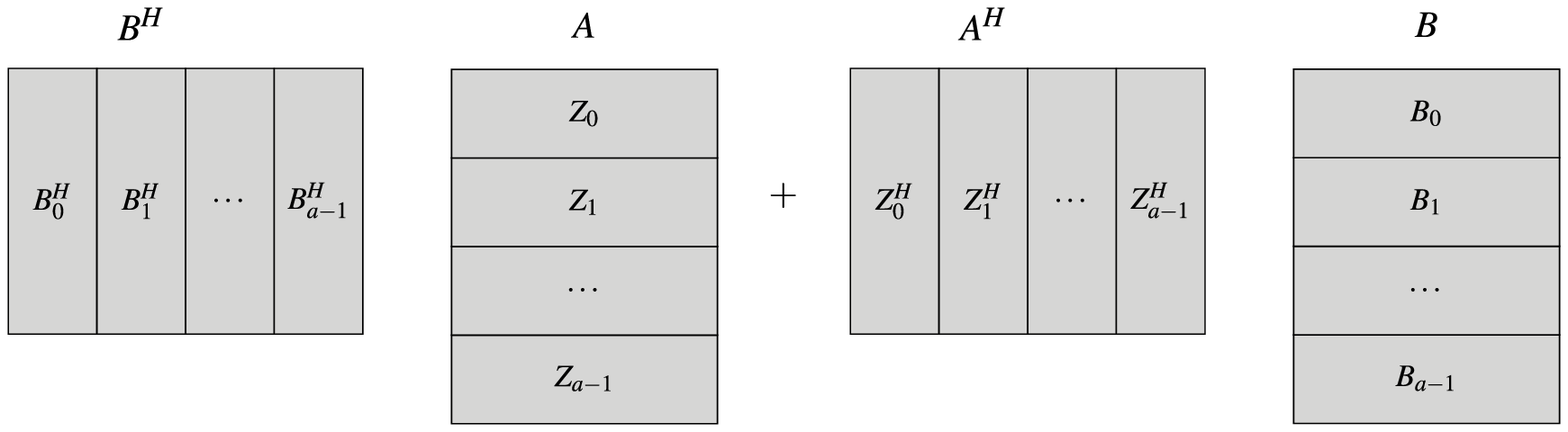}
    \caption{Stacking of multiple small matrices $Z_a$ and $B_a$ into larger 
    matrices $A$ and $B$, respectively. As a result, multiple small matrix
    products are combined into a larger better performing product.}
    \label{fig:agg}
\end{figure}

For implementation purposes, the computation of $H$ is split into two parts, $H_{AB+BA+BB}$ and  $H_{AA}$ (Eq.~\ref{eq:h}),
which can be cast completely in terms of BLAS-3 operations. 
The key idea behind the calculation of $H_{AB+BA+BB}$ (lines 3--9) is to
rewrite the expression as 
$$\sum^{N_A}_{a=1} B^H_a (T^{[BA]} A_a) + (A^H_a T^{[AB]}) B_a + \frac{1}{2} B^H_a (T^{[BB]} B_a) + \frac{1}{2} (B^H_a T^{[BB]}) B_a,$$
where $T^{[BA]}$ is the Hermitian transpose of $T^{[AB]}$, and
$T^{[BB]}$ is Hermitian. By doing so, the common expressions in parentheses can be grouped together and substituted with
\[
 Z_a = T^{[BA]} A_a + \frac{1}{2} T^{[BB]} B_a,
\]
for each atom $a$ (lines 3--5). The matrices $Z_a$ and $B_a$ are then stacked
together into larger matrices $A$ and $B$ (lines 6--7), allowing the remaining computation
$$ \sum^{N_A}_{a=1} B^H_a Z_a + Z_a^H B_a $$
to be performed as one single large matrix product (line 9), as depicted in Fig.~\ref{fig:agg}.

To compute the second part of the matrix $H$ (lines 16--28), the algorithm again takes advantage of the existing symmetries in $H_{AA}$.
Since the $T^{[AA]}$ matrix is Hermitian positive definite (HPD), HSDLA first attempts to compute the Cholesky factorization of $T^{[AA]}$ ($C_a C_a^H =
T^{[AA]}$) (line 18), which is then substituted in $H_{AA}$ yielding

\[
 H_{AA} = \sum^{N_A}_{a=1} A^H_a T^{[AA]} A_a \quad \Rightarrow \quad \sum^{N_A}_{a=1} (\underbrace{C_a^H A_a}_{Z_a})^H (C_a^H A_a).
\]

However, while in theory $T^{[AA]}$ is HPD, in practice, due to numerical considerations, the
factorization may fail. The algorithm thus divides the computation of $H_{AA}$
in two parts.
In case the factorization succeeds, the matrix $Z_a$ is computed (line 20) and stacked at
the top part of $B$ ($B_T$) (line 21).
If, on the other hand, the factorization fails, the matrix
$Z_a$ (line 23) is stacked at the bottom of matrix $B$ ($B_B$), and the matrix
$A_a$ stacked in $A$.
At the completion of the loop, $H$ is updated with the operation $B^H_{T} B_{T}
+ A^H B_{B}$ (lines 27--28), where the first term ($B^H_{T} B_{T}$) exploits
the symmetry and computes only half of the output via the BLAS routine {\tt
herk}, while the second term ($A^H B_{B}$) is computed via the BLAS routine
{\tt gemm}, which computes a full matrix.

Finally, the computation of $S$ (lines 12--14) is more straightforward. First,
the product $A^H A$ is computed as one single large product. Then $B$ is updated
with the norms stored in $U$ and then second large product $B^H B$ completes the
computation of $S$.

\section{Related work} \label{sec:relwork}

Today, a numerous optimized BLAS-like computational libraries exist.
However, up to our knowledge, none of these libraries does not implement
BLAS kernels capable for automatic redistribution of the computational flops 
between different computational units, such as CPU, GPU and/or Phi.
The most popular multi-GPU commercial BLAS library, CUBLASXT,
supports hybrid CPU-GPU computation, but so far, only {\tt gemm}
kernels are implemented as hybrid. The bottom-right part of the input matrix
is offloaded to the CPU and the percentage (i.e. the number of rows/columns) 
that will be offload to CPU is set at the compile time. 
An academic alternative to CUBLASXT, called
BLASX~\cite{blasx} requires minor changes to the calls to the BLAS routines
and, similar to CUBLASXT library, takes care of data transfers between
CPU and GPU transparently. Although the authors reported a significant
speedup and the communication volume decrease compared to CUBLASXT,
Magma and some other libraries and runtimes, again, only the {\tt dgemm}
routine is implemented in the hybrid fashion, i.e. with a possibility to
schedule execution between CPUs and GPUs. The other kernels of
interest for this work have CPU-only and multi-GPU-only
implementations. 

Further research in hybridization of codes has been done in~\cite{hcyltkd14} 
in which authors developed
a lightweight scheduler to offload tasks between different computational
units inside a single compute node. The scheduler is based on task 
parallelism and, as was demonstrated in the case of Cholesky, the 
factorization tasks are executed on the CPU while updating is done 
on the GPUs. The authors also demonstrated the portability and the
scalability of the their approach across different accelerators,
including multi-GPU and multi-Phi systems. However, there is no report
on hybrid BLAS-3 kernels required by HSDLA. 

At the time of writing this paper, only one research~\cite{SolcaKHTDS15} 
reported on hybrid GPU-based algorithms for the generation of Hamiltonian and Overlap matrices 
in FLAPW methods. In that work, the authors aimed at 1000+ atoms systems, too
large to be executed on a single compute node thus a distributed block-cyclic 
setup and distribution of the Hamiltonian and Overlap matrices were implemented.
However, the authors based the generation of $H$ and $S$ only on distributed versions of
{\tt zgemm} and did not exploit symmetries to decrease computational
cost. In addition, the generation is a task-based heterogeneous
implementation where particular tasks are scheduled between CPU and
the GPUs. In our approach to heterogeneity, we aim at
further exploiting data parallelism, not task parallelism.

\section{HSDLA on Heterogeneous Architectures} \label{sec:hdslaHetero}

In this section, we concentrate on the behaviour of the algorithm when executed
on hybrid architectures consisting of shared-memory CPUs and one or more
graphic processor units (GPUs) or Intel Xeon Phi coprocessors.  As discussed
in~\cite{DiNapoli:HSDLA}, once HSDLA is cast in terms of BLAS and LAPACK
routines, it attains high performance on multi-core architectures by simply
linking to a multi-threaded implementation of these libraries. However, as we
demonstrate in this section, attaining such performance levels on hybrid architectures 
is not straightforward. 

We first refine the original HSDLA algorithm, and significantly reduce its
computational cost as well as its memory footprint. 
Then, we identify the limitations of a straightforward port to hybrid
architectures and the shortcomings of the existing solutions, and study
alternative approaches to attain satisfactory performance levels on such
architectures.

\subsection{HSDLA refined} \label{ssec:refined}

In the effort of exploiting symmetries from the problem, the designers of HSDLA
overlooked the fact that matrix $H$ is Hermitian and used the {\tt zgemm} routine
(general matrix-matrix product) in Algorithm~\ref{alg:hands} (line 28) instead of
a routine that takes advantage of the symmetry and computes only one triangle of
the output matrix. As a result, the algorithm performs $4_{\neg{}HPD}N_A N_L N_G^2$ redundant
flops for computing the upper triangle of the matrix. 
With that in mind, and given that some vendor libraries such as NVIDIA's CUBLASXT
and Intel's MKL provide a specialized routine to compute only one half of a general
matrix product, line 16 through 28 can be greatly simplified, avoiding the Cholesky
factorization altogether, by replacing them with the computation in Algorithm~\ref{alg:refinement}.
As a result, the total cost of computing $H_{AA}$ is significantly reduced and is 
dominated by the specialized routine with a cost of $4 N_A N_L N_G^2$.

\begin{algorithm}
\begin{algorithmic}[1]
\makeatletter
\setcounter{ALG@line}{14}
\makeatother
    \State
    // $H_{AA}$ 
    \For{$a := 1 \to N_A$}
        \State $X_a = T^{[AA]}_a A_a$          \Comment{({\tt zhemm}: $8 N^2_L N_G$ Flops)}
        \State Stack $X_a$ to $X$
        \State Stack $A_a$ to $A$
    \EndFor
    \State $H = H + A^H X$  \Comment{({\tt zherkx}: $4 N_A N_L N^2_G$ Flops)}
\end{algorithmic}
    \caption{{\bf: Refinement of the computation of $H_{AA}$.}}
    \label{alg:refinement}
\end{algorithm}

By further examining Algorithm~\ref{alg:hands} one observes two sequential
steps, a backup and a restore of matrices $A$ and $B$, lines 1 and 11,
respectively. Since $A$ and $B$ are overwritten (lines 3--9), one copy of each
are stored in two additional storage spaces $\hat{A}$ and $\hat{B}$ (line 1)
and reused to compute $S$ (lines 12--14) and the rest of $H$ ($H_{AA}$) (lines
16--28). However, the need for additional storage spaces $\hat{A}$ and $\hat{B}$ can be
significantly reduced by simply reordering the execution flow as described in
Algorithm~\ref{alg:hdslaNew}. The computation of $S$, which only requires one
temporary buffer, is moved to the beginning of the algorithm (lines 2--4); the
result of computing $UB$ is stored in the temporary buffer $X$, and the original
values $A$ and $B$ and are so far preserved.
The buffer $X$ is then reused as auxiliary storage in the computation of
$H_{AB+BA+BB}$ for stacking $Z_a$ (line 9); note that overwriting $B$ now in
line 10 is harmless since its original contents are not needed anymore. 
Finally, in the computation of $H_{AA}$, the result of line 15 is also stacked
in $X$, and $A$ is compressed in the $A$ buffer itself.
Since all of $A$, $B$, and $X$ are of size $16 N_{A} N_{L} N_{G}$ bytes (in the
order of gigabytes), cutting this requirements in half has a major impact on
the memory footprint of the entire algorithm.

\begin{algorithm}[!h]
\begin{algorithmic}[1]
    \\// S
    \State $S = A^H A$      \Comment{({\tt zherk}: $4 N_A N_L N^2_G$ Flops)}
    \State $X = U B$        \Comment{({\tt scaling}: $2 N_A N_L N_G$ Flops)}
    \State $S = S + X^H X$  \Comment{({\tt zherk}: $4 N_A N_L N^2_G$ Flops)}
    
    \\// $H_{AB + BA + BB}$ 
    \For{$a := 1 \to N_A$}
        \State $Z_a = T^{[BA]}_a A_a$  \Comment{({\tt zgemm}: $8 N^2_L N_G$ Flops)}
        \State $Z_a = Z_a + \frac{1}{2} T^{[BB]}_a B_a$  \Comment{({\tt zhemm}: $8 N^2_L N_G$ Flops)}
        \State Stack $Z_a$ to $X$
        \State Stack $B_a$ to $B$
    \EndFor
    \State $H = X^H B + B^H X$      \Comment{({\tt zher2k}: $8 N_A N_L N^2_G$ Flops)} 
    
    \\// $H_{AA}$ 
    \For{$a := 1 \to N_A$}
        \State $X_a = T^{[AA]}_a A_a$          \Comment{({\tt zhemm}: $8 N^2_L N_G$ Flops)}
        \State Stack $X_a$ to $X$
        \State Stack $A_a$ to $A$
    \EndFor
    \State $H = H + A^H X$  \Comment{({\tt zherkx}: $4 N_A N_L N^2_G$ Flops)}
    
\end{algorithmic}
    \caption{{\bf: The refined HSDLA without Cholesky and with only one temporary buffer.}}
    \label{alg:hdslaNew}
\end{algorithm}

\subsection{Limitations of the straightforward approach} \label{ssec:limitations}

As we reported in our previous work~\cite{jhpcs/Fabregat}, with very
limited effort, HSDLA can be easily ported to various computing
architectures such as multi-core and multi-GPU systems. To achieve
that goal, wrappers have to be implemented around calls to optimized
architecture-specific libraries such as cuBLAS, CUBLASXT or Intel MKL,
and function calls changed with those of the optimized libraries for a
particular architecture. Although this straightforward approach can
quickly bring reasonable performance, it cannot utilize the underlying
computational system at their full potential. For example, if CUBLASXT
is used on a multi-GPU system, most of the BLAS operations will be
offloaded to the GPUs while the CPUs remain idle.  Thus, simply using
the existing optimized libraries cannot yield the performance increase
that is expected by combining the power of all CPUs and GPUs of a
system.

In order to overcome this issue, we developed heterogeneous BLAS kernels for
the routines {\tt zher2k}, {\tt zherk}, and {\tt zgemm/zherkx} that can efficiently divide
the workload between the CPUs and the accelerators (GPUs and Phis).
In the rest of this section, we describe two different designs on how to
implement the BLAS kernels on heterogeneous architectures, required by the
refined HDSLA. We denoted these two approaches as:

\begin{itemize}
	\item {\it Static} - 
	The computation between the CPUs and GPUs is divided by pre-computing the number of rows/columns that will be offloaded to the GPUs and then a highly-tuned multi-GPU library is used to compute it.
	\item {\it Dynamic} -
	The computation is based on dynamic scheduling depending on available on-device memory to determine block size and per-device task queuing. 
\end{itemize}

\subsection{Static}\label{ssec:static}

This approach targets hybrid architectures with one or more CUDA-based GPU
devices, but can be easily extended to Intel Xeon Phi as well as other
accelerators, as long as optimized BLAS libraries for these architectures
exist.  The key idea of this approach is to re-use the existing, highly
optimized multi-GPU and multi-threaded libraries in building the hybrid code,
with the goal to significantly reduce the programming effort. As such, the code
can be quickly tuned for new and emerging GPU architectures, thus improving
performance portability on different platforms, while at the same time explores
concurrent execution on all available CPUs and GPUs of the system.

For the sake of simplicity, only the {\tt zher2k} routine will be
described in details, while the same approach can be easily extended
to the other two routines.  The {\tt zher2k} performs the Hermitian
rank 2k update of the given matrix $C$ in one of the following
operations:
\begin{align*}
    C &:= \alpha A B^H + \overline{\alpha} B A^H + \beta C,
    \\ 
    C &:= \alpha A^H B + \overline{\alpha} B^H A + \beta C,
\end{align*}
where $\alpha$ and $\beta$ are scalars, $C\in R^{n\times n}$ is a Hermitian
matrix, and $A$ and $B$ are general matrices of size ${n\times k}$ in the first
case and $k\times n$ in the second case.  Hereafter, we will observe only the
first case; the same algorithm applies for the second, but with $A$
and $B$ of transposed dimensions.

\begin{figure}
    \centering
    \includegraphics[width=0.85\textwidth]{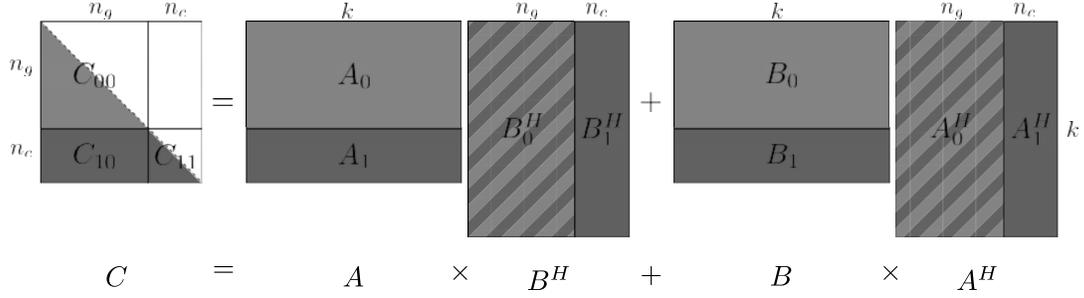}
    \caption{Rank 2k update of matrix $C$ ({\tt zher2k}). The blocks are statically split between the CPUs (dark grey) and GPUs (light grey). $n_g$ is the size of principal submatrix offloaded to the GPUs.}
    \label{fig:static}
\end{figure}

Matrix $C$ is split into four blocks and updated as described in Figure~\ref{fig:static}. The workload is divided such that the principal submatrix $C_{00}$ 
with dimension $n_g \times n_g$ is updated on the GPU(s), while the blocks $C_{10}$ and $C_{11}$ ($C_{01}$ is not required since $C$ is Hermitian) are updated on the CPU. 
The entire update of $C$ can then be performed with only 4 function calls:
\begin{align*}
 C_{00} &= \alpha A_0 B_0^H + \overline{\alpha} B_0 A_0^H + \beta C_{00} &\longrightarrow \text{zher2k (GPU)}\\
 C_{11} &= \alpha A_1 B_1^H + \overline{\alpha} B_1 A_1^H + \beta C_{11} &\longrightarrow \text{zher2k (CPU)}\\
 C_{10} &= \overline{\alpha} B_1 A_0^H + \beta C_{10} &\longrightarrow \text{zgemm (CPU)} \\
 C_{10} &= \alpha A_1 B_0^H + C_{10} &\longrightarrow \text{zgemm (CPU)}.
\end{align*}

Since these operations are independent from each other, they can be performed in
parallel. For scheduling the workload between CPU and GPU devices, OpenMP with
nested parallelism is used. Two OpenMP threads are created, one manages the computation
on the CPU and invokes multi-threaded MKL kernels, the other manages the computation
on the GPU and invokes CUBLASXT kernels on the GPU.

Although this approach is simple, the main challenge that remains is how to
optimally set the size of the principal submatrix that will be offloaded to the
GPU. In the ideal case the CPU and GPU execution times should be completely
overlapped. Thus, knowing that the total number of flops for {\tt zher2k} is $8
k n^2$, with $n$ the number of rows and columns of $C$ and $k$ number of
columns of matrices $A$ and $B$, $n_g$, the size of the principal submatrix
$C_{00}$, may be computed as:

\begin{align*}
  \text{time}_{gpu} &= \text{time}_{cpu} \\[5pt]
  \frac{\text{flops}_{gpu}}{\text{Gflops}_{gpu}} &= \frac{\text{flops}_{cpu}}{\text{Gflops}_{cpu}} \\[5pt]
  \frac{8 k n_g^2 + 4}{\text{Gflops}_{gpu}} & = \frac{8 k (n-n_g)^2 + 4 + 16 k n_g (n-n_g)}{\text{Gflops}_{cpu}}
\end{align*}

Note that the updating block $C_{10}$ requires two additional calls to {\tt zgemm} which adds extra flops to the count.
If $m = \frac{\text{Gflops}_{gpu}}{\text{Gflops}_{cpu}}$ then the size of matrix offloaded to the GPU is:
\[
 n_g^2 = \frac{m n^2 + 4 m}{m + 1}.
\]

As we illustrate in Section~\ref{sec:experiments}, the optimal offload between CPU and GPU devices 
can be computed analytically using this approach or the performance ratio between CPU and GPU can be estimated on 
smaller problem sizes.

\subsection{Dynamic}\label{ssec:dynamic}

While a static split of the matrices and a static assignment of blocks to
devices leads to the lowest overheads and highest efficiency, it lacks the
flexibility to adapt easily to new circumstances.  An alternative is the use of
a dynamic scheduling that queries the devices to determine the available
on-device memory, calculates appropriate block sizes, and uses per-device
queues to distribute the work.
To maximize portability, the implementation is split into a
device-specific and a device-independent part.  This allows the
dynamic code to target both the Xeon Phi coprocessor and CUDA GPUs.
The device-independent part contains all the workload distribution,
scheduling, the BLAS operation and hybrid calculation logic.  As for the
device-dependent code, the Xeon Phi implementation builds on the
hStreams library and MKL, while the GPU implementation relies on CUDA
and cuBLAS.

The device-independent logic works as follows. First, each matrix is split into
square blocks.  The block size is chosen to split the matrices evenly, fit
multiple blocks into GPU memory, and follow vendor recommendations (e.g. for
the first generation Xeon Phi accelerators the block size should be divisible by 64 but not 256).
Next, the scheduler determines the individual operations to be performed on each of
the submatrices, and schedules them to the devices in a round-robin fashion.
Each block is packed into contiguous buffers and streamed to the devices; the 
results are unpacked once the calculation and the reverse transfer complete.
The packing of the blocks is necessary for the Xeon Phi, since it lacks support 
for copies from 2D arrays, but it is avoided on GPUs.
If the scheduler notices that a devices is not yet ready to accept work because its queue
is full, it skips that devices.
Once the memory of the devices are filled up, and the CPU is not needed to drive
the devices, the CPU starts computing block operations using its own computational
resources. 

The remaining challenge is to improve utilization at the very start of a
calculation, and at the very end.  There, it is desirable to reduce block
sizes and sacrifice kernel efficiency for load balance.  However, implementing
such a scheme requires sophisticated models of transfer and compute efficiency,
and it is left to future work.

\section{Experimental Results} \label{sec:experiments}

We turn now our attention to experimental results.  We compare the
performance of the \emph{original} multi-core (CPU only) HSDLA
algorithm \cite{DiNapoli:HSDLA} against our \emph{refined} CPU
implementation from Section~\ref{ssec:refined} and the two hybrid
CPU-GPU implementations based on that refined algorithm.  As mentioned
in Section~\ref{sec:hdslaHetero} these include an implementation that
offloads large matrix-matrix products to the multi-GPU using
\emph{CUBLASXT}, an implementation using \emph{static} work assignment
(Section~\ref{ssec:static}) and an implementation using \emph{dynamic}
work assignment (Section~\ref{ssec:dynamic}).  Since the
implementation using dynamic work assignment supports offloading to
Intel Xeon Phi accelerators, we also present results on that platform.

As test cases we use three input systems that describe distinct
physical systems.  We refer to them as NaCl, AuAg and TiO$_2$,
respectively.  These systems represent a heterogeneous sample
(including both insulators and conductors) with different physical
properties.  In our tests, the code generates the matrices $H$ and $S$
for one single \kv-point, and different $K_{max}$ values.  The actual
problem sizes, that is, the values for $N_A$, $N_L$, and $N_G$ for
each case are given in Tab.~\ref{tab:sizes}.

\begin{table}[!ht]
    \centering
    \begin{tabular}{ c  |  c  c  |  c  c  c  c  c } \hline
        {\bf Test case} & $N_A$ & $N_L$ & $N_G:$ & 
                          {\scriptsize $K_{max}=2.5$} & 
                          {\scriptsize $K_{max}=3.0$} & 
                          {\scriptsize $K_{max}=3.5$} & 
                          {\scriptsize $K_{max}=4.0$} \\ \hline
        NaCl      & 512 &  49 & & 2\,256 & \ \ 3\,893 & \ \ 6\,217 & \ \  9\,273 \\
        AuAg      & 108 & 121 & & 3\,275 & \ \ 5\,638 & \ \ 8\,970 & 13\,379 \\
        TiO$_2$      & 384 &  81 & & 7\,094 & 12\,293 & 19\,553 & 29\,144 \\\hline
    \end{tabular}
    \caption{Problem sizes for NaCl, AuAg, and TiO$_2$, for a variety of $K_{max}$ values.
             The value of $N_G$ varies with $K_{max}$.}
    \label{tab:sizes}
\end{table}

\subsection{Experimental setup}

We ran our experiments on a range of compute nodes.  We used two
computed nodes hosted by the IT center of the RWTH Aachen University.
One of these nodes (RWTH-GPU) consists of two eight-core Sandy Bridge
E5-2680 processors, running at a frequency of 2.7 GHz. The node is
equipped with 64 GBs of RAM and 2 Nvidia Tesla K20Xm GPUs.  The peak
performance of the 16 CPU cores in double precision is 345
GFlops/s, while the peak performance of each GPU is 1.3 TFlops/s,
for a combined peak of almost 3.0 TFlops/s.  The second RWTH node
(RWTH-Phi) consists of two eight-core Sandy Bridge E5-2650 processors,
running at a frequency of 2.0 GHz. The node is also equipped with 64
GBs of RAM and 2 Intel Xeon Phi 5110p accelerators (Knight's Corner),
The peak performance of the 16 CPU cores in double precision is 256
GFlops/s, while the peak performance of each Xeon Phi is (about) 1
TFlops/s, for a combined peak of 2.3 TFlops/s.

We also ran experiments on the JURECA supercomputer at the J\"ulich
Supercomputing Centre (JSC). More specifically, we used one JURECA node
consisting of two twelve-core Haswell E5-2680v3 processors,
running at a nominal frequency of 2.5 GHz, and 2 NVIDIA K80 GPUs (each of which
consists of two GK210 devices). The node is equipped with 128 GBs of RAM.
The combined peak performance for the 24 CPU cores in double precision is 
960 GFlops/s, while the theoretical peak performance in double precision of each GPU device
is about 1.45 TFlops/s, for a total of 6.7 TFlops/s.

In the following subsections, when presenting efficiency of the
achieved system with a varying number of accelerators, we take as
reference value the peak performance corresponding to the same number
of accelerators. For example, when presenting the results on JURECA
with 1, 2 and 3 GPUs, the system peak performance is 2.4, 3.8, and 5.3
TFlops, respectively.

We compare several different improved HSDLA versions.
We refer to the base HSDLA algorithm, as implemented in~\cite{DiNapoli:HSDLA}, with {\it original}.
Our algorithmically refined multiCPU HSDLA version
is denoted as {\it refined}, while the same version but for multi-GPU system is named {\it cuBlasXt}.
The accelerated versions, denoted as {\it static} and {\it dynamic},
are the version described in Sections~\ref{ssec:static} and~\ref{ssec:dynamic}, respectively.
In all cases, the code was linked to Intel MKL version 11.3.2 for the BLAS
routines on the CPU and Xeon Phis; the GPU code was linked to NVIDIA
CUBLASXT version 7.5 and 8.0 on the RWTH-GPU nodes and on the JURECA nodes, respectively.

\subsection{RWTH-GPU}

\begin{figure}
\centering
\includegraphics[width=.8\textwidth]{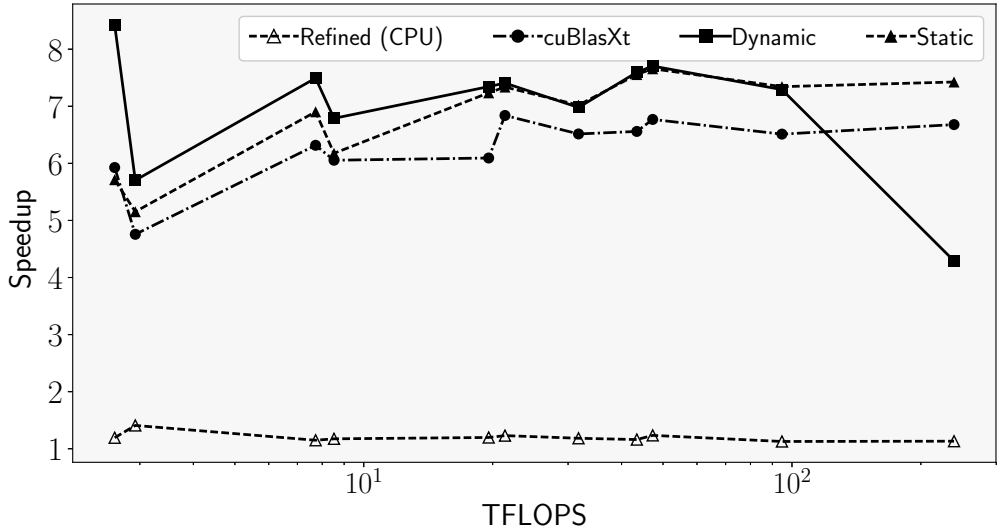}
\caption{Speedup on RWTH-GPU for all implementations, relative to
  original code.}\label{fig:rwth-gpu-speedup}
\end{figure}


\begin{figure}
\centering
\includegraphics[width=.8\textwidth]{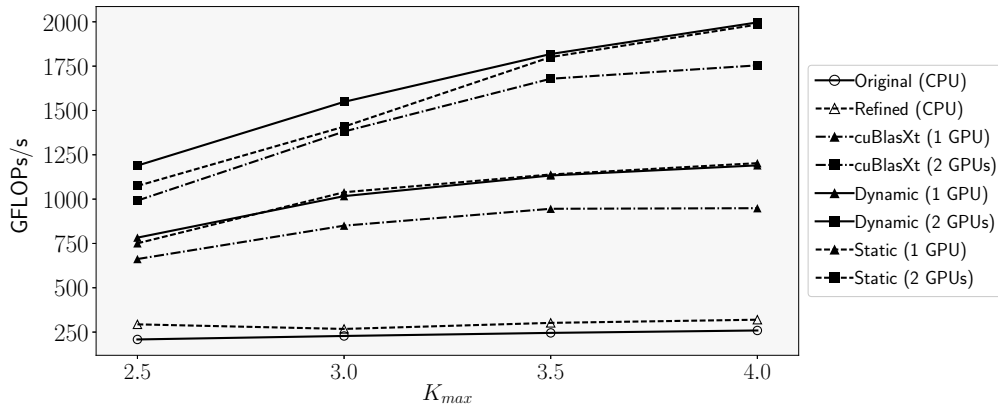}
\caption{Performance on RWTH-GPU for all implementations, for the AuAg test case.}\label{fig:rwth-gpu-perf}
\end{figure}


We provide two experiments for the RWTH-GPU system.
%
Figure~\ref{fig:rwth-gpu-speedup} presents the speedup over the original HSDLA
algorithm attained by the refined, cuBlasXt, dynamic, and static
implementations, for each of the different test cases 
ordered by the amount of computation required by each of them.
%
The results show an average speedup of $1.13$ of the refined algorithm
over the original, when run on the CPU only.  That clearly
demonstrates how small changes in the code, such as exploiting
symmetries and reordering of the execution flow (requiring limited
programming effort), can lead to higher performance, even without
using the accelerators.

Also, the figure illustrates how the GPU code (cuBlasXt, static, and
dynamic) achieves seizable speedups no matter how large the
computation is.  In most cases our custom hybrid implementations
outperform the cuBlasXT, thanks to both better tuning of block sizes
and especially a heavier usage of the host CPU computational power.
The dip at the end for the dynamic code is due to TiO$_2$ case run out of
memory when 2 GPUs are used and instead result using 1 GPU is presented.

Figure~\ref{fig:rwth-gpu-perf} showcases the total performance achieved by the
different implementations for the AuAg test case and a range of values for
$K_{max}$.  Again, the advantages of enabling the use of the entire compute
nodes (CPU + GPUs) become apparent. There is a clear separation between
CPU-only and accelerated code, and also between GPU-only cuBlasXt and the
hybrid static and dynamic codes. 
Both our hybrid codes outperform GPU-only (cuBlasXt), for large enough
problems, by more than $200$ Gflops, which is very close to the peak
performance of the CPU-only version, Table~\ref{tab:rz-auag}. This clearly demonstrates
that the all CPUs are fully utilized in our hybrid approach.
Furthermore, for large problems that allow for a reasonable utilization
of the multiple GPUs, we achieve about 2 TFlops/s compared to a peak performance of 3 TFlops/s,
 i.e. 66\% efficiency. The efficiency is even higher for larger problems
such as TiO$_2$ with $K_{max} = 4.0$, where our code attains about 2.5 TFlops/s (83\%
of the peak). 

The presented results include timings from the
CPU-only parts of the code (Algorithm~\ref{alg:hdslaNew} lines 6--11 and 14--18), during which the 
GPUs are in idle state, thus decreasing total performance and efficiency. The efficiency 
attained by our hybrid BLAS routines is even higher, as presented in Table~\ref{tab:rz-auag}.
The attained efficiency of the BLAS-only kernels is up to 79\% for AuAg case and slightly better for
TiO$_2$ case, up to 81\% of the peak system performance.

\begin{table}[ht]
	\caption{GFlops/s and efficiency (in parentheses) for the AuAg test case and $K_{max} = 4.0$.
	Results for the RWTH node using 2 GPUs.}
    \centering
    \begin{tabular}{ l | c  c  c  } 

                     \hline
        {\bf Implementation} & {\bf zherk} & {\bf zher2k} & {\bf zherkx/zgemmt} \\ 
                      \hline
        CPU        &   358 (11.93\%) &  304 (10.13\%) &  343 (11.43\%)  \\
        cuBlasXt   &  2041 (68.30\%) & 2012 (67.06\%) & 2021 (67.07\%) \\
        Dynamic    &  2332 (77.73\%) & 2367 (78.90\%) & 2302 (76.73\%) \\
        Static     &  2337 (77.91\%) & 2315 (77.16\%) & 2315 (77.16\%) \\
        \hline
    \end{tabular}
    \label{tab:rz-auag}
\end{table}

\subsection{RWTH-Phi}

\begin{figure}[h]
\centering
\includegraphics[width=.8\textwidth]{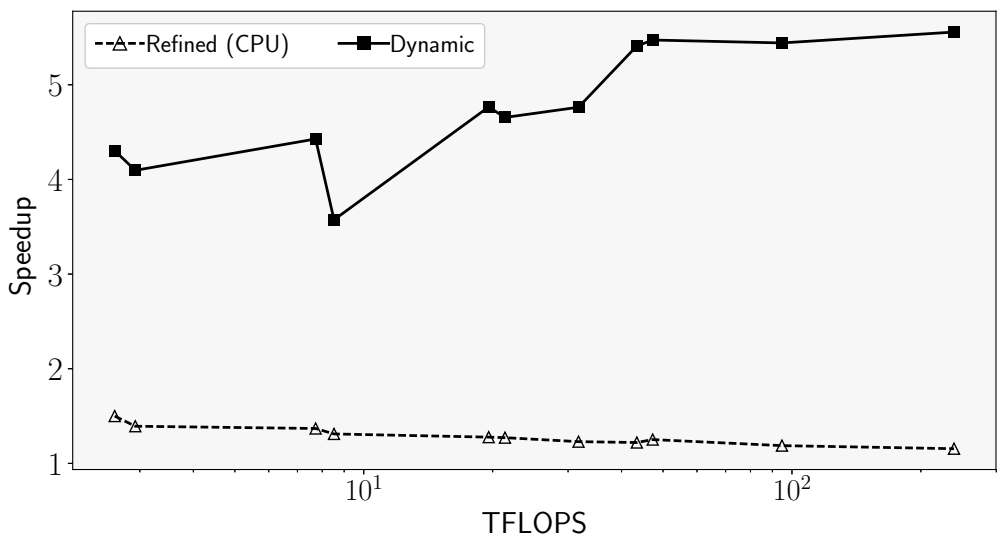}
\caption{Speedup on RWTH-Phi for the refined and the hybrid dynamic implementations, relative to original code.}\label{fig:rwth-phi-speedup}
\end{figure}

\begin{figure}[h]
\centering
\includegraphics[width=.8\textwidth]{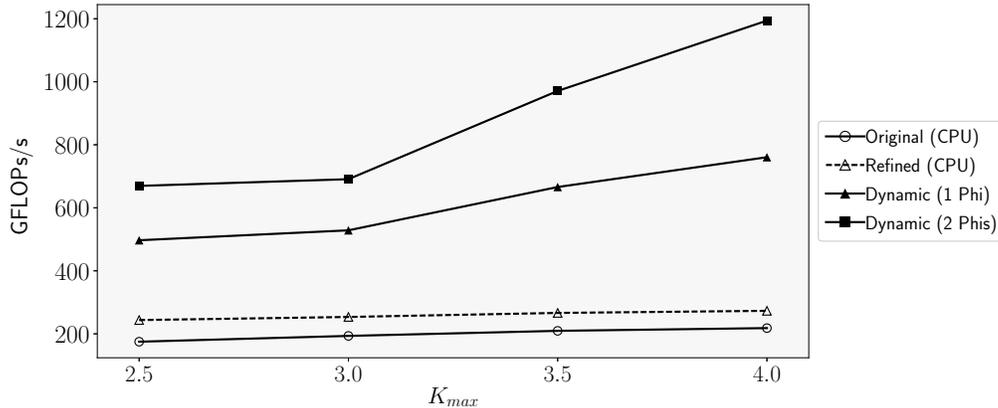}
\caption{Performance on RWTH-Phi for the original, refined and dynamic implementations, for the AuAg test case.}\label{fig:rwth-phi-perf}
\end{figure}

For the RWTH-Phi system, we again present a speedup plot
(Fig.~\ref{fig:rwth-phi-speedup}) and a performance plot
(Fig.~\ref{fig:rwth-phi-perf}).  In this case, we present results for the
refined algorithm and the dynamic version of the hybrid code, which is the only
one with current support for the Xeon Phi (through the hStreams API).
While MKL can offload calculations automatically to the device, this
is only true for the GEMM operation. Therefore, the automatic offload
is not applicable to our algorithm, which is dominated by the {\tt
  zherk}, {\tt zher2k} and {\tt zherkx} routines.

Figure~\ref{fig:rwth-phi-speedup} shows how our hybrid code
consistently achieves a speedup between 4 and 5 with respect to the
original HSDLA algorithm.  This is consistent with both the relative
power of the two accelerator cards and the CPU, as well as the
expectation that the Phi accelerators do not get as close to their
theoretical peak performance as GPUs do.

In Figure~\ref{fig:rwth-phi-perf}, we illustrate the performance of
the original, refined, and dynamic codes for AuAg test case. For the
larger case ($K_{max}$ = 4.0), the dynamic versions attain about 700
GFlops/s, using one Phi, and 1200 GFlops/s, using two Phis, which
correspond to 55\% and 52\% of the system peak performance,
respectively. For the TiO$_2$ test case we observed even higher
performances of up to 950 GFlops/s (one Phi) and 1600 GFlops/s (two
Phis) corresponding to an efficiency of 75\% and 69\%.
Similarly to the previous experiments in the RWTH-GPU system, the
larger the problem size, the larger the efficiency attained, which
shows that even higher efficiency is to be expected when larger
systems are simulated.

\subsection{JURECA-GPU}

\begin{figure}[h]
\centering
\includegraphics[width=.8\textwidth]{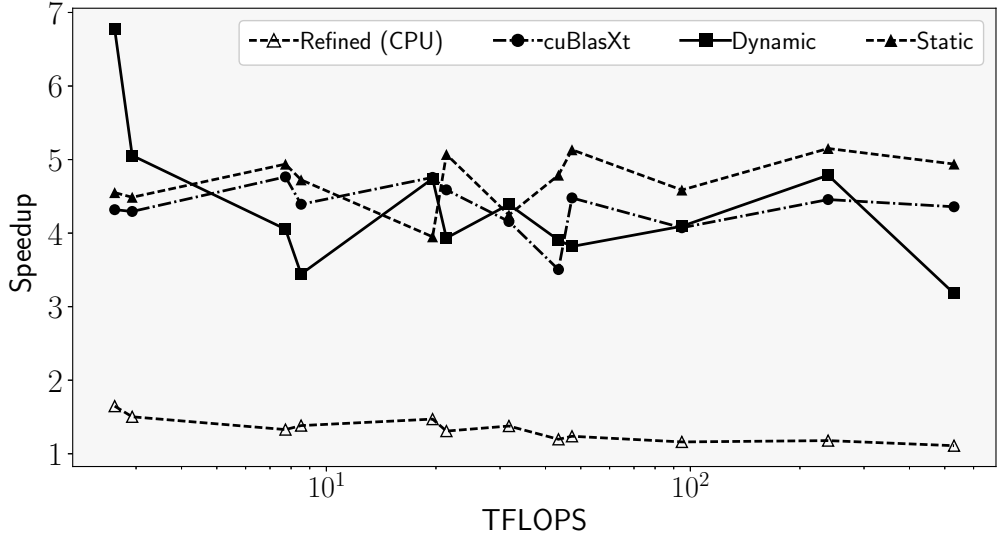}
\caption{Speedup on JURECA for all implementations, relative to original code.}\label{fig:jureca-speedup}
\end{figure}

\begin{figure}[h]
\centering
\includegraphics[width=.8\textwidth]{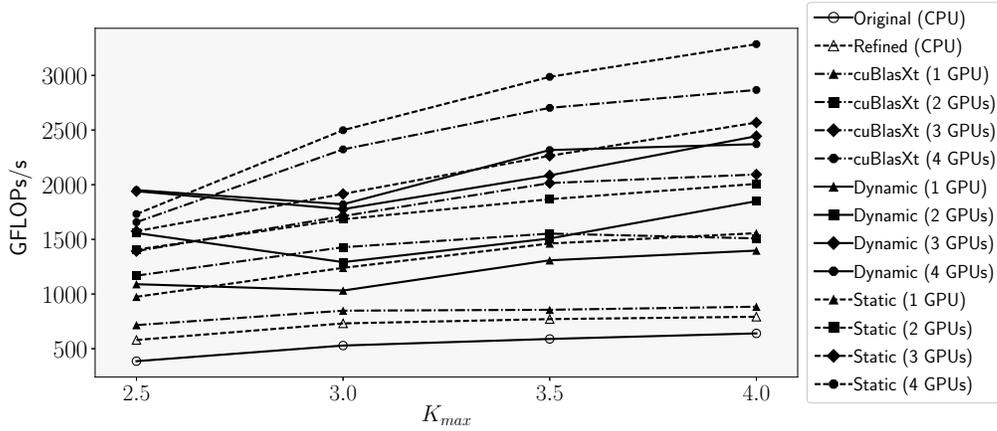}
\caption{Performance on JURECA for all implementations, for the AuAg test case.}\label{fig:jureca-perf}
\end{figure}

\begin{figure}[h]
\centering
\includegraphics[width=.8\textwidth]{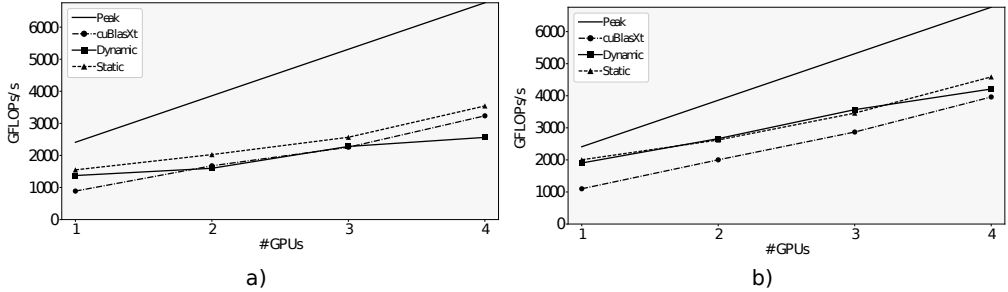}
\caption{Scalability of the hybrid BLAS routines on JURECA going from one to many GPUs, for both codes, for the AuAg a) and TiO$_2$ b) 3.5 test case.}
\label{fig:jureca-scale}
\end{figure}

The JURECA-GPU nodes comprise two NVIDIA K80 GPUs, each of which
consists of two devices.  This allows us not only to not evaluate
speedups (Fig.~\ref{fig:jureca-speedup}) and performance
(Fig.~\ref{fig:jureca-perf}), but also scalability of the different
implementations (Fig.~\ref{fig:jureca-scale}).
Figure~\ref{fig:jureca-speedup} illustrates the speedup of the various
GPU implementations relative to the original HSDLA. The speedup ranges
most of the times between 3.5 and 5.5 (starting from a maximum of about 7).
The fact that the speedup of the dynamic code varies widely indicates
that the scheduler struggles to fully utilize all four GPUs.
With regards to the comparison among the three GPU-hybrid implementations,
there is again a clear separation between the static code and the
cuBlasXt implementation.

The performance plot in Fig.~\ref{fig:jureca-perf} illustrates the
flop rates achieved by each implementation.  The clear separation
between the static and the cuBlasXt implementation is again very
pronounced.  However, as observed in Fig.~\ref{fig:jureca-speedup},
the dynamic implementation struggles to utilize many GPUs.  It seems
like further improvements would be necessary to properly adapt it to
this use case.
Overall, our static implementation achieves up to 3.2 TFlops/s for the
AuAg test case, that is, a 15\% higher performance than the cuBlasXt
implementation.  Similar results are achieved for the TiO$_2$ case, in
which we observe a performance of up to 4.3 TFlops/s on 4 GPUs,
resulting in 16\% higher performance compared to the cuBlasXt
implementation.
As the system has a peak performance of 6.7 TFlops/s (4 GPU included),
the results for AuAg and TiO$_2$ cases correspond to 47\% and 61\%
peak performance utilization, respectively. These much lower results
compared to RWTH-GPU are the result of the surprisingly low performance
of the CUBLASXT implementations (only 55\% utilization) of the key
BLAS-3 kernels, on which our accelerated HSDLA versions depend on.

The plot in Fig.~\ref{fig:jureca-scale} captures the scalability
behaviour of our code for the AuAg and TiO$_2$, $K_{max}$ = 3.5 case,
for 1, 2, 3, and 4 GPUs. In this plot we focus solely on our hybrid
implementation of the off-loaded BLAS routines, and consider only
this part of the runtime.
We extract three important messages from the graph: First, the
measurements for static and dynamic code nearly coincide.  This is not
surprising, since both codes aim to achieve the same goal: hybrid
execution.
Second, the measurements for all three implementations roughly share
the same slope.  This indicates that all codes utilize the GPUs
similarly well as additional devices are added.  Third, there is a gap
between the hybrid codes and the GPU-only code of almost 1 TFlop/s,
which is roughly the computational power of the CPU. This reflects
that not only the hybrid codes reach a utilization of the GPUs similar
to the vendor optimized CUBLASXT, but they also fully utilize the CPU
cores.
This is particularly of interest because modern CPUs provide high computational 
power (e.g.~960 GFlops on JURECA per node), even when compared to that of the 
coprocessors, and thus should not be neglected nor underutilized. Table~\ref{tab:jureca-tio2} 
shows that the performance gain of combining CPUs with GPUs in the BLAS-3 kernels 
in the HSDLA algorithm can improve performance up to 19\% compared to the state-of-the-art
cuBlasXt implementations using 4 GPUs.
  
\begin{table}[ht]
	\caption{GFlops/s and speedup (in parentheses) for {\it static} and {\it dynamic} implementations compared to {\it cuBlasXt} for the TiO$_2$ test case and $K_{max} = 3.5$ on JURECA using 4 GPUs.}
    \centering
    \begin{tabular}{ l | l  l  l  } 
        \hline
        {\bf Implementation} & {\bf zherk} & {\bf zher2k} & {\bf zherkx/zgemmt} \\ 
                      \hline
        cuBlasXt   &  3439 \phantom{(5.72$\times$)} & 3258 \phantom{(6.64$\times$)} & 3257 \phantom{(5.89$\times$)} \\
        Dynamic    &  3862 (1.12$\times$) & 3590 (1.11$\times$) & 3266 (1.01$\times$) \\
        Static     &  4012 (1.19$\times$) & 3832 (1.19$\times$) & 3686 (1.13$\times$) \\
        \hline
    \end{tabular}
    \label{tab:jureca-tio2}
\end{table}

\section{Conclusions and Future Work} \label{sec:conclusions}

In this paper we demonstrated how legacy codes, such as FLEUR code,
can be modernized in order to exploit the massive parallelism of
modern computing architectures, and to improve code performance,
portability and scalability.  The starting point of our work was the
HSDLA algorithm. HSDLA encodes the computation of the Hamiltonian and
Overlap matrices ($H$ and $S$), one of the two computational
bottlenecks of the FLEUR code, in terms of multi-dimensional
operations that map well onto BLAS operations. However, the
straightforward porting of these BLAS operations to heterogeneous
architectures consisting of multi-core CPU and one or more GPUs or Phi
accelerators, does not attain expected performances. The reason behind
the limited performance is that vendor optimized BLAS libraries do not
provide support for such hybrid architectures and exploit either the
CPUs or the accelerators, but not both.

We used the original BLAS-based HSDLA code as a starting point to
explore both existing and custom implementations in order to take
advantage of the accelerators.  First, we improved the original HSDLA
algorithm by introducing changes in the execution flow; the result was
a much lower memory footprint, and a reduced computational cost.
Then, we showed that custom hybrid implementations can boost
performance compared to accelerator-only implementations, especially
when CPUs themselves provide large computational power.

We presented two approaches to implement hybrid BLAS: A dynamic
approach that schedules chunks of calculation on the devices using
buffers and queues, and a static approach that splits matrices based
on prescribed ratios. The dynamic approach is more generic, as
demonstrated by targeting both GPUs and Phis.  Because the static code
needs to be tuned through the ratios, it gives better control for
tuning.  Both implementation strategies considerably boost performance
due to their hybrid nature.

However, there is still room for improvement. By using mathematical
equalities the objects $A$ and $B$ can be compressed into smaller
objects, leading to less but more complex computations.  More
specifically, the compressed form of $A$ and $B$ leads to tensor
contractions that do not naturally map onto {\tt gemm}-like operations
and require the development of specialized routines. If efficient
mappings and implementations of new kernels are found, speedups of up
to one order of magnitude may be achieved. Furthermore, in order to
solve even larger problems (systems of up to 1000 atoms), the
presented approaches present certain limitations such as the memory
requirements and the limited size of the on-device memory. To
efficiently solve larger problems new memory and
communication-avoiding approaches on distributed multi-CPU and
multi-GPU systems are required, including the redesign of the most
time-consuming kernels.

\section*{Acknowledgements}
This work was partially funded by 
the Ministry of Science and Education of the Republic of Croatia and
the Deutsche Akademische Austauschdienst (DAAD) from funds of the 
Bundesministeriums f\"ur Bildung und Forschung (BMBF) through project 
``PPP Kroatien'' ID 57216700.
Financial support from the J\"ulich Aachen Research Alliance-High Performance
Computing and the Deutsche Forschungsgemeinschaft (DFG) through grant GSC 111 is
also gratefully acknowledged. Furthermore, the authors thank the RWTH IT Center
and the J\"ulich Supercomputing Centre for the computational
resources.

\bibliographystyle{acm}
\bibliography{biblio}

\end{document}